\documentclass[12pt, draftcls, onecolumn]{IEEEtran}
\usepackage{epsfig, psfrag}
\newtheorem{thm}{Theorem}[section]

\newcommand{\Input}{\mathbf{s}}
\newcommand{\Output}{\mathbf{x}}

\newcommand{\cFading}{\mathbf{h}_\mathrm{c}}
\newcommand{\Fading}{\mathbf{h}_\mathrm{d}}

\newcommand{\Noise}{\mathbf{z}}
\newcommand{\Innovation}{\mathbf{v}}

\newcommand{\cspectral}{S_{\cFading}}
\newcommand{\spectral}{S_{\Fading}}

\newcommand{\SNR}{\rho}
\newcommand{\MSE}{\epsilon}
\newcommand{\ErrorVar}{\sigma^2}

\newcommand{\ctranspose}{^\dag}
\newcommand{\expect}{\mathcal{E}}

\newcommand{\define}{\stackrel{\mathrm{def}}{=}}

\newcommand{\complex}{\mathcal{C}}
\newcommand{\cGaussian}{\mathcal{CN}}
\renewcommand{\baselinestretch}{1.5}

\begin{document}

\title{\LARGE{How Good is Phase-Shift Keying for Peak-Limited Rayleigh Fading Channels in the Low-SNR Regime?}}
\author{\normalsize{Wenyi Zhang, {\it Student Member, IEEE}, and J.~Nicholas Laneman, {\it Member, IEEE}}
\thanks{This work has been supported in part by the State of Indiana through the 21st Century Research Fund, by the National Science Foundation through contract ECS03-29766, and by the Fellowship of the Center for Applied Mathematics of the University of Notre Dame. Some preliminary results in this work were presented in part at the IEEE International Workshop on Signal Processing Advances for Wireless Communication (SPAWC), New York, 2005.}
\thanks{The authors are with Department of Electrical Engineering, University of Notre Dame. Email: {\tt \{wzhang1,jnl\}@nd.edu}}
}

\maketitle

\begin{abstract}
This paper investigates the achievable information rate of phase-shift keying (PSK) over frequency non-selective Rayleigh fading channels without channel state information (CSI). The fading process exhibits general temporal correlation characterized by its spectral density function. We consider both discrete-time and continuous-time channels, and find their asymptotics at low signal-to-noise ratio (SNR). Compared to known capacity upper bounds under peak constraints, these asymptotics usually lead to negligible rate loss in the low-SNR regime for slowly time-varying fading channels. We further specialize to case studies of Gauss-Markov and Clarke's fading models.
\end{abstract}

\section{Introduction}
\label{sec:intro}

For Rayleigh fading channels without channel state information (CSI) at low signal-to-noise ratio (SNR), the capacity-achieving input gradually tends to bursts of ``on'' intervals sporadically inserted into the ``off'' background, even under vanishing peak power constraints \cite{sethuraman05:it}. This highly unbalanced input usually imposes implementation challenges. For example, it is difficult to maintain carrier frequency and symbol timing during the long ``off'' periods. Furthermore, the unbalanced input is incompatible with linear codes, unless appropriate symbol mapping ({\it e.g.}, $M$-ary orthogonal modulation with appropriately chosen constellation size $M$) is employed to match the input distribution.

This paper investigates the achievable information rate of phase-shift keying (PSK). PSK is appealing because it has constant envelope and is amenable to linear codes without additional symbol mappings. Focusing on low signal-to-noise ratio (SNR) asymptotics, we utilize a recursive training scheme to convert the original fading channel without CSI into a series of parallel sub-channels, each with estimated CSI but additional noise that remains circular complex white Gaussian. The central results in this paper are as follows. First, for a discrete-time channel whose unit-variance fading process $\left\{\Fading[k]: -\infty < k < \infty \right\}$ has a spectral density function $\spectral(e^{j\Omega})$ for $-\pi \leq \Omega \leq \pi$, the achievable rate is $(1/2)\cdot\left[ (1/2\pi)\cdot\int_{-\pi}^\pi \spectral^2(e^{j\Omega})d\Omega - 1 \right]\cdot \SNR^2 + o(\SNR^2)$ nats per symbol, as the average channel SNR $\SNR \rightarrow 0$. This achievable rate is at most $(1/2)\cdot \SNR^2 + o(\SNR^2)$ away from the channel capacity under peak SNR constraint $\SNR$. Second, for a continuous-time channel whose unit-variance fading process $\left\{\cFading(t): -\infty < t < \infty\right\}$ has a spectral density function $\cspectral(j\omega)$ for $-\infty < \omega < \infty$, the achievable rate as the input symbol duration $T \rightarrow 0$ is $\left[1 - (1/2\pi P)\cdot\int_{-\infty}^\infty \log\left(1 + P \cdot \cspectral(j\omega)\right)d\omega\right]\cdot P$ nats per unit time, where $P > 0$ is the envelope power. This achievable rate exactly coincides with the channel capacity under peak envelope $P$.

We further apply the above results to specific case studies of Gauss-Markov fading models (both discrete-time and continuous-time) as well as a continuous-time Clarke's fading model. For discrete-time Gauss-Markov fading processes with innovation rate $\MSE \ll 1$, the quadratic behavior of the achievable rate becomes dominant only for $\SNR \ll \MSE$. Our results, combined with previous results for the high-SNR asymptotics, suggest that coherent communication can essentially be realized for $\MSE \leq \SNR \leq 1/\MSE$. For Clarke's model, we find that the achievable rate scales sub-linearly, but super-quadratically, as $O\left(\log(1/P)\cdot P^2\right)$ nats per unit time as $P \rightarrow 0$.

The remainder of this paper is organized as follows. Section \ref{sec:model} describes the channel model and the recursive training scheme. Section \ref{sec:asymptotic} deals with the discrete-time channel model, and Section \ref{sec:bandwidth} the continuous-time channel model. Finally Section \ref{sec:conclusion} provides some concluding remarks. Throughout the paper, random variables are in bold font. All the logarithms are to base $e$, and information units measured in nats.

\section{Channel Model, Recursive Training Scheme, and Effective SNR}
\label{sec:model}

We consider a scalar time-selective, frequency non-selective Rayleigh fading channel, written in baseband-equivalent continuous-time form as
\begin{eqnarray}
\label{eqn:channel-ct}
\Output(t) = \cFading(t) \cdot \Input(t) + \Noise(t),\quad\mathrm{for}\; -\infty < t < \infty,
\end{eqnarray}
where $\Input(t) \in \complex$ and $\Output(t) \in \complex$ denote the channel input and the corresponding output at time instant $t$, respectively. The additive noise $\{\Noise(t): -\infty < t < \infty\}$ is modeled as a zero-mean circular complex Gaussian white noise process with $\expect\{\Noise(s) \Noise\ctranspose(t)\} = \delta(s - t)$. The fading process $\{\cFading(t): -\infty < t < \infty\}$ is modeled as a wide-sense stationary and ergodic zero-mean circular complex Gaussian process with unit variance $\expect\{\cFading(t) \cFading\ctranspose(t)\} = 1$ and with spectral density function $\cspectral(j\omega)$ for $-\infty < \omega < \infty$. Additionally, we impose a technical condition that $\{\cFading(t): -\infty < t < \infty\}$ is mean-square continuous, so that its autocorrelation function $K_{\cFading}(\tau) = \expect\{\cFading(t + \tau)\cFading\ctranspose(t)\}$ is continuous for $\tau \in (-\infty, \infty)$.

Throughout the paper, we restrict our attention to PSK over the continuous-time channel (\ref{eqn:channel-ct}). For technical convenience, we let the channel input $\Input(t)$ have constant envelope $P > 0$ and piecewise constant phase, {\it i.e.},
\begin{eqnarray*}
\Input(t) = \Input[k] = \sqrt{P} \cdot e^{j\theta[k]},\quad\mathrm{if}\;kT \leq t < (k + 1)T,
\end{eqnarray*}
for $-\infty < k < \infty$\footnote{Here we note a slight abuse of notation in this paper, that a symbol ({\it e.g.}, $\Input$) can be either continuous-time or discrete-time. The two cases are distinguished by $\cdot(t)$ for continuous-time and $\cdot[k]$ for discrete-time.}. The symbol duration $T > 0$ is determined by the reciprocal of the channel input bandwidth\footnote{For multipath fading channels, $T$ should also be substantially greater than the delay spread \cite{biglieri98:it}, otherwise the frequency non-selective channel model (\ref{eqn:channel-ct}) may not be valid. Throughout the paper we assume that this requirement is met.}.

Applying the above channel input to the continuous-time channel (\ref{eqn:channel-ct}), and processing the channel output through a matched filter\footnote{A matched filter suffers no information loss for white Gaussian channels \cite{proakis95:book}. For the fading channel (\ref{eqn:channel-ct}), it is no longer optimal in general \cite{kailath60:it}. However, in this paper we still focus on the matched filter, which is common in most practical systems.}, we obtain a discrete-time channel
\begin{eqnarray}
\label{eqn:channel-dt}
\Output[k] = \sqrt{\SNR} \cdot \Fading[k] \cdot \Input[k] + \Noise[k],\quad\mathrm{for}\; -\infty < k < \infty.
\end{eqnarray}
The channel equations (\ref{eqn:channel-ct}) and (\ref{eqn:channel-dt}) are related through
\begin{eqnarray*}
\Input[k] &=& e^{j \theta[k]}\\
\Output[k] &=& \frac{1}{\sqrt{T}} \int_{kT}^{(k + 1)T} \Output(t) dt\\
\Noise[k] &=& \frac{1}{\sqrt{T}} \int_{kT}^{(k + 1)T} \Noise(t) dt\\
\Fading[k] &=& \frac{1}{\sqrt{\int_0^T \int_0^T K_{\cFading}(s - t) ds dt}} \int_{kT}^{(k + 1)T} \cFading(t) dt.
\end{eqnarray*}

For the discrete-time channel (\ref{eqn:channel-dt}) we can verify that
\begin{itemize}
\item The additive noise $\{\Noise[k]: -\infty < k < \infty\}$ is circular complex Gaussian with zero mean and unit variance, {\it i.e.}, $\Noise[k] \sim \cGaussian(0, 1)$, and is independent, identically distributed (i.i.d.) for different $k$.
\item The fading process $\{\Fading[k]: -\infty < k < \infty\}$ is wide-sense stationary and ergodic zero-mean circular complex Gaussian, with $\Fading[k]$ being marginally $\cGaussian(0, 1)$. We further notice that $\{\Fading[k]: -\infty < k < \infty\}$ is obtained through sampling the output of the matched filter, hence its spectral density function is
\begin{eqnarray*}
\spectral(e^{j\Omega}) = \frac{1}{\sqrt{\int_0^T \int_0^T K_{\cFading}(s - t) ds dt}} \cdot \sum_{k = -\infty}^\infty \cspectral\left(j\frac{\Omega - 2 k \pi}{T}\right) \cdot \mathrm{sinc}^2(\Omega - 2 k \pi)
\end{eqnarray*}
for $-\pi \leq \Omega \leq \pi$.
\item The channel input $\{\Input[k]: -\infty < k < \infty\}$ is always on the unit circle. In the sequel, we will further restrict it to be complex proper \cite{neeser93:it}, {\it i.e.}, $\expect\{\Input^2[k]\} = \left [\expect\{\Input[k]\}\right ]^2$. The simplest such input is quadrature phase-shift keying (QPSK); by contrast, binary phase-shift keying (BPSK) is not complex proper.
\item The average channel SNR is given by
\begin{eqnarray}
\label{eqn:SNR}
\SNR &=& \frac{P}{T} \cdot \left (\int_0^T \int_0^T K_{\cFading}(s - t) ds dt\right ) > 0.
\end{eqnarray}
\end{itemize}

Throughout the paper, we assume that the realization of the fading process $\{\cFading(t): -\infty < t < \infty\}$ is not directly available to the transmitter or the receiver, but its statistical characterization in terms of $\cspectral(j\omega)$ is precisely known at both the transmitter and the receiver.

\begin{figure}[tbp]
\psfrag{CH0}{{$\mathrm{PSC}_0$}}
\psfrag{CH1}{{$\mathrm{PSC}_1$}}
\psfrag{CHL}{{$\mathrm{PSC}_{L - 1}$}}
\psfrag{Lsubchannels}{$L$ parallel sub-channels (PSC)}
\psfrag{Klengthblocks}{length-$K$ coding blocks}
\epsfxsize=3.3in
\epsfclipon
\centerline{\epsffile{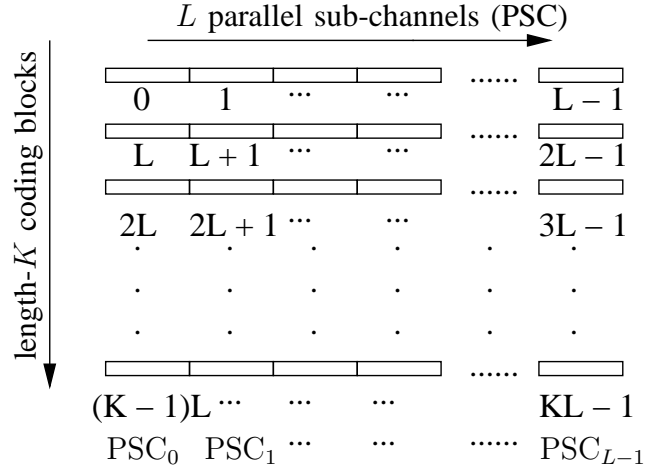}}
\caption{Illustration of the interleaving scheme. Input symbols are encoded/decoded column-wise, and transmitted/received row-wise.}
\label{fig:interleave}
\end{figure}

We employ a recursive training scheme to communicate over the discrete-time channel (\ref{eqn:channel-dt}). By interleaving the transmitted symbols as illustrated in Figure \ref{fig:interleave} (cf. \cite{goldsmith96:it}), the recursive training scheme effectively converts the original non-coherent channel into a series of parallel sub-channels, each with estimated receive CSI but additional noise that remains i.i.d. circular complex Gaussian. The interleaving scheme decomposes the channel into $L$ parallel sub-channels (PSC). The $l$th ($l = 0, 1, \ldots, L - 1$) PSC sees all the inputs $\Input[k \cdot L + l]$ of (\ref{eqn:channel-dt}) for $k = 0, 1, \ldots, K - 1$. These $L$ PSCs suffer correlated fading, and this correlation is exactly what we seek to exploit using recursive training. Although some residual correlation remains within each PSC among its $K$ symbols, due to the ergodicity of the channel (\ref{eqn:channel-dt}), this correlation vanishes as the interleaving depth $L \rightarrow \infty$. In practical systems with finite $L$, if necessary, we may utilize an additional interleaver for each PSC to make it essentially memoryless.

We make a slight abuse of notation in the sequel. Since all the PSCs are viewed as memoryless, when describing a PSC we can simply suppress the internal index $k$ among its $K$ coding symbols, and only indicate the PSC index $l$ without loss of generality. For example, $\Fading[l]$ actually corresponds to any $\Fading[k \cdot L + l]$, for $k = 0, 1, \ldots, K - 1$.

The recursive training scheme performs channel estimation and demodulation/decoding in an alternating manner. To initialize transmission, PSC 0, the first parallel sub-channel, transmits pilots rather than information symbols to the receiver. Based upon the received pilots in PSC 0, the receiver predicts $\Fading[1]$, the fading coefficient of PSC 1, and proceeds to demodulate and decode the transmitted symbols in PSC 1 coherently. If the rate of PSC 1 does not exceed the corresponding channel mutual information, then information theory ensures that, as the coding block length $K \rightarrow \infty$, there always exist codes that have arbitrarily small decoding error probability. Hence the receiver can, at least in principle, form an error-free reconstruction of the transmitted symbols in PSC 1, which then effectively become ``fresh'' pilots to facilitate the prediction of $\Fading[2]$ and subsequent coherent demodulation/decoding of PSC 2. Alternating the estimation-demodulation/decoding procedure repeatedly, all the PSCs are reliably decoded one after another.

{\it Remark:} A major drawback of the recursive training scheme is that its interleaved structure typically leads to a large delay. The coding block length $K$ should be large enough such that the decoding error probability is small enough to prevent catastrophic error propagation along the PSCs. Furthermore, the number of PSCs $L$ should also be large enough such that the prediction of the fading process essentially converges to its steady-state limit. Only after receiving all the $K \cdot L$ symbols in the interleaved block can the receiver perform the alternating estimation-demodulation/decoding procedure. However, we note that this may not be the case for wideband channels. In wideband channels with frequency-decorrelated fading processes, we can employ multi-carrier techniques, {\it e.g.}, orthogonal frequency-division multiplexing (OFDM), to decompose the original wide bandwidth into a large number of sub-bands, suffering essentially independent frequency non-selective fading processes. In this case, each row in Figure \ref{fig:interleave} corresponds to a sub-band, and the coding block length ({\it i.e.}, the number of sub-bands) $K$ increases as the bandwidth grows. For each PSC, its $K$ coding symbols occur simultaneously in physical time, hence the receiver need not wait until receiving all the $K \cdot L$ symbols to perform the alternating estimation-demodulation/decoding procedure.

By induction, let us consider PSC $l$, assuming that the inputs $\{\Input[i]: i = 0, 1, \ldots, l - 1\}$ of the previous PSCs have all been successfully reconstructed at the receiver. Since the channel inputs are always on the unit circle, the receiver can compensate for their phases in the channel outputs, and the resulting observations become
\begin{eqnarray*}
\underbrace{e^{-j\theta[i]} \cdot \Output[i]}_{\Output^\prime[i]} = \sqrt{\SNR} \cdot \Fading[i] + \underbrace{e^{-j\theta[i]} \cdot \Noise[i]}_{\Noise^\prime[i]}\quad\mathrm{for}\;i = 0, 1, \ldots, l - 1.
\end{eqnarray*}
Since zero-mean circular complex Gaussian distributions are invariant under rotation, the rotated noise $\Noise^\prime[i]$ is still i.i.d. zero-mean unit-variance circular complex Gaussian. Then we can utilize standard linear prediction theory ({\it e.g.}, \cite{kailath81:book}) to obtain the one-step minimum mean-square error (MMSE) prediction of $\Fading[l]$ defined as
\begin{eqnarray}
\hat{\Fading}[l] = \expect\left\{\Fading[l] \left.\right| \left\{\Output^\prime[i]: i = 0, 1, \ldots, l - 1\right\} \right\}.
\end{eqnarray}
The estimate $\hat{\Fading}[l]$ and the estimation error $\tilde{\Fading}[l] = \Fading[l] - \hat{\Fading}[l]$ are jointly circular complex Gaussian distributed as $\cGaussian\left (0, 1 - \ErrorVar[l]\right )$ and $\cGaussian\left (0, \ErrorVar[l]\right )$, respectively, and are uncorrelated and further independent. Here $\ErrorVar[l]$ denotes the mean-square prediction error. The channel equation of PSC $l$ can then be written as
\begin{eqnarray}
\Output[l] &=& \sqrt{\SNR} \cdot \Fading[l] \cdot \Input[l] + \Noise[l]\nonumber\\
\label{eqn:channel-coh}
&=& \sqrt{\SNR} \cdot \hat{\Fading}[l] \cdot \Input[l] + \underbrace{\sqrt{\SNR} \cdot \tilde{\Fading}[l] \cdot \Input[l] + \Noise[l]}_{\bar{\Noise}[l]},
\end{eqnarray}
where the effective noise $\bar{\Noise}[l]$ is circular complex Gaussian, and is independent of both the channel input $\Input[l]$ and the estimated fading coefficient $\hat{\Fading}[l]$. Thus, the channel (\ref{eqn:channel-coh}) becomes a coherent Gaussian channel with fading and receive CSI $\hat{\Fading}[l]$, with effective SNR
\begin{eqnarray}
\label{eqn:snr-l}
\SNR[l] = \frac{1 - \ErrorVar[l]}{\ErrorVar[l]\cdot\SNR + 1} \cdot \SNR.
\end{eqnarray}

In the paper we mainly focus on the ultimate performance limit without delay constraints, which is achieved as the interleaving depth $L \rightarrow \infty$. Under mild technical conditions, the one-step MMSE prediction error sequence $\{\ErrorVar[l]: l = 0, 1, \ldots\}$ converges to the limit \cite[Chap. XII, Sec. 4]{doob53:book}
\begin{eqnarray}
\label{eqn:errorvar-inf}
\ErrorVar_\infty \define \lim_{l \rightarrow \infty} \ErrorVar[l] = \frac{1}{\SNR} \cdot \left\{\exp\left\{ \frac{1}{2\pi} \int_{-\pi}^\pi \log\left ( 1 + \SNR \cdot \spectral(e^{j\Omega}) \right ) d\Omega \right\} - 1\right\}.
\end{eqnarray}
Consequently the effective SNR (\ref{eqn:snr-l}) sequence $\{\SNR[l]: l = 0, 1, \ldots\}$ converges to
\begin{eqnarray}
\label{eqn:snr-inf}
\SNR_\infty \define \lim_{l \rightarrow \infty} \SNR[l] = \frac{1 - \ErrorVar_\infty}{\ErrorVar_\infty \cdot \SNR + 1} \cdot \SNR.
\end{eqnarray}

We are mainly interested in evaluating the mutual information of the induced channel (\ref{eqn:channel-coh}) at the limiting effective SNR $\SNR_\infty$ as the actual channel SNR $\SNR \rightarrow 0$. This low-SNR channel analysis is facilitated by the explicit second-order expansion formulas of the channel mutual information at low SNR \cite{prelov04:it}. Applying \cite[Theorem 3]{prelov04:it} to the induced channel\footnote{Note that \cite[Theorem 3]{prelov04:it} is only applicable to complex proper channel inputs, as we have assumed in the channel model.} (\ref{eqn:channel-coh}) at $\SNR_\infty$, we have
\begin{eqnarray}
\label{eqn:rate}
R \define \lim_{l \rightarrow \infty} I\left(\Input[l]; \Output[l] \left.\right| \hat{\Fading}[l]\right) &=& \SNR_\infty - \SNR_\infty^2 + o(\SNR^2) \quad\mathrm{as}\;\SNR \rightarrow 0.
\end{eqnarray}

\section{Asymptotic Channel Mutual Information at Low SNR}
\label{sec:asymptotic}

As shown in (\ref{eqn:rate}), the asymptotic channel mutual information depends on the limiting effective SNR (\ref{eqn:snr-inf}), which further relates to the limiting one-step MMSE prediction error (\ref{eqn:errorvar-inf}). The following theorem evaluates the asymptotic behavior of the channel mutual information (\ref{eqn:rate}).
\begin{thm}
\label{thm:minfo-expand}
For the discrete-time channel (\ref{eqn:channel-dt}), as $\SNR \rightarrow 0$, its induced channel (\ref{eqn:channel-coh}) achieves the rate
\begin{eqnarray}
\label{eqn:minfo-expand}
R = \frac{1}{2} \left [ \frac{1}{2\pi} \int_{-\pi}^\pi \spectral^2(e^{j\Omega}) d\Omega - 1 \right ] \cdot \SNR^2 + o(\SNR^2),
\end{eqnarray}
if the integral $(1/2\pi) \cdot \int_{-\pi}^\pi \spectral^2(e^{j\Omega}) d\Omega$ exists.
\end{thm}
{\it Proof:} We will prove that
\begin{eqnarray}
\label{eqn:errorvar-expand}
\ErrorVar_\infty = 1 - \frac{1}{2} \left [ \frac{1}{2\pi} \int_{-\pi}^\pi \spectral^2(e^{j\Omega}) d\Omega - 1\right ] \cdot \SNR + o(\SNR),
\end{eqnarray}
which together with (\ref{eqn:snr-inf}) leads to
\begin{eqnarray*}
\SNR_\infty = \frac{1}{2} \left [ \frac{1}{2\pi} \int_{-\pi}^\pi \spectral^2(e^{j\Omega}) d\Omega - 1\right ] \cdot \SNR^2 + o(\SNR^2).
\end{eqnarray*}
Then (\ref{eqn:minfo-expand}) immediately follows from (\ref{eqn:rate}).

For simplicity let us denote by $g(\SNR)$ the integral $(1/2\pi) \cdot \int_{-\pi}^\pi \log \left ( 1 + \SNR \cdot \spectral(e^{j\Omega}) \right ) d\Omega$, hence
\begin{eqnarray*}
\lim_{\SNR \rightarrow 0} g(\SNR) &=& \frac{1}{2\pi} \int_{-\pi}^\pi \log 1 d\Omega = 0\\
\lim_{\SNR \rightarrow 0} \frac{dg(\SNR)}{d\SNR} &=& \frac{1}{2\pi} \int_{-\pi}^\pi \spectral(e^{j\Omega})d\Omega = 1\\
\lim_{\SNR \rightarrow 0} \frac{dg^2(\SNR)}{d^2\SNR} &=& -\frac{1}{2\pi} \int_{-\pi}^\pi \spectral^2(e^{j\Omega})d\Omega.
\end{eqnarray*}
To prove (\ref{eqn:errorvar-expand}), we apply l'Hospital's rule in (\ref{eqn:errorvar-inf}) to evaluate
\begin{eqnarray*}
\lim_{\SNR \rightarrow 0} \ErrorVar_\infty &=& \lim_{\SNR \rightarrow 0} \frac{e^{g(\SNR)} - 1}{\SNR}\nonumber\\
&=& \lim_{\SNR \rightarrow 0} e^{g(\SNR)}\cdot \frac{dg(\SNR)}{d\SNR} = 1;\\
\lim_{\SNR \rightarrow 0} \frac{d(\ErrorVar_\infty)}{d\SNR} &=& \lim_{\SNR \rightarrow 0} \left [ \frac{e^{g(\SNR)}}{\SNR}\cdot \frac{dg(\SNR)}{d\SNR} - \frac{e^{g(\SNR)} - 1}{\SNR^2}\right ]\\
&=& \lim_{\SNR \rightarrow 0} \left [ e^{g(\SNR)}\cdot \left ( \frac{dg(\SNR)}{d\SNR} \right )^2 + e^{g(\SNR)} \cdot \frac{dg^2(\SNR)}{d^2\SNR} - \frac{e^{g(\SNR)}}{2\SNR} \cdot \frac{dg(\SNR)}{d\SNR} \right ]\\
&=& \frac{1}{2} \lim_{\SNR \rightarrow 0} \left [ e^{g(\SNR)}\cdot \left ( \frac{dg(\SNR)}{d\SNR} \right )^2 + e^{g(\SNR)} \cdot \frac{dg^2(\SNR)}{d^2\SNR} \right ]\\
&=& - \frac{1}{2} \left [ \frac{1}{2\pi}\int_{-\pi}^\pi \spectral^2(e^{j\Omega})d\Omega - 1 \right ].
\end{eqnarray*}
Substituting the above quantities into the first-order Taylor expansion of $\ErrorVar_\infty$, we then obtain (\ref{eqn:errorvar-expand}).   {\bf Q.E.D.}

Theorem \ref{thm:minfo-expand} states that for PSK at low SNR, the achievable channel mutual information vanishes quadratically with SNR. This is consistent with \cite{medard02:it} \cite{hajek02:it}. Furthermore, it is of particular interest to compare the asymptotic expansion (\ref{eqn:minfo-expand}) with several previous results.

\subsection{Comparison with a Capacity Upper Bound}
For the discrete-time channel (\ref{eqn:channel-dt}), PSK with constant SNR $\SNR$ is a particular peak-limited channel input. The capacity per unit energy of channel (\ref{eqn:channel-dt}) under a peak SNR constraint $\SNR$ is \cite{sethuraman05:it}
\begin{eqnarray}
\label{eqn:capacity-dt}
\dot{C} = 1 - \frac{1}{2\pi\SNR} \cdot \int_{-\pi}^\pi \log \left ( 1 + \SNR \cdot \spectral(e^{j\Omega}) \right )d\Omega,
\end{eqnarray}
achieved by on-off keying (OOK) in which each ``on'' or ``off'' symbol corresponds to an infinite number of channel uses, and the probability of choosing ``on'' symbols vanishes. Such ``bursty'' channel inputs are in sharp contrast to PSK. From (\ref{eqn:capacity-dt}), an upper bound to the channel capacity can be derived as \cite{sethuraman05:it}
\begin{eqnarray}
\label{eqn:capacity-ub-dt}
C \leq U(\SNR) \define \frac{1}{2} \cdot \frac{1}{2\pi} \int_{-\pi}^\pi \spectral^2(e^{j\Omega}) d\Omega \cdot \SNR^2.
\end{eqnarray}
Comparing (\ref{eqn:minfo-expand}) and (\ref{eqn:capacity-ub-dt}), we notice that the penalty of using PSK instead of the bursty capacity-achieving channel input is at most $(1/2)\cdot\SNR^2 + o(\SNR^2)$. For fast time-varying fading processes, this penalty can be relatively significant. For instance, if the fading process is memoryless, {\it i.e.}, $\spectral(e^{j\Omega}) = 1$ for $-\pi \leq \Omega \leq \pi$, then $(1/2\pi) \cdot \int_{-\pi}^\pi \spectral^2(e^{j\Omega})d\Omega - 1 = 0$, implying that no information can be transmitted using PSK over a memoryless fading channel. Fortunately, for slowly time-varying fading processes, the integral $(1/2\pi) \cdot \int_{-\pi}^\pi \spectral^2(e^{j\Omega})d\Omega$ is typically much greater than $1$, as we will illustrate in the sequel.

\subsection{Comparison with the High-SNR Channel Behavior}
From (\ref{eqn:minfo-expand}) and (\ref{eqn:capacity-ub-dt}), it can be said that $(1/2\pi) \cdot \int_{-\pi}^\pi \spectral^2(e^{j\Omega})d\Omega$ is a fundamental quantity associated with a fading process at low SNR. This is in contrast to the high-SNR regime, where a fundamental quantity is \cite{lapidoth05:it}
\begin{eqnarray*}
\ErrorVar_\mathrm{pred} \define \exp\left\{ \frac{1}{2\pi}\int_{-\pi}^\pi \log\spectral(e^{j\Omega}) d\Omega \right\}.
\end{eqnarray*}
The quantity $\ErrorVar_\mathrm{pred}$ is the one-step MMSE prediction error of $\Fading[0]$ given its entire noiseless past $\{\Fading[-1], \Fading[-2], \ldots\}$. When $\ErrorVar_\mathrm{pred} > 0$ the process is said to be regular; and when $\ErrorVar_\mathrm{pred} = 0$ it is said to be deterministic, that is, the entire future $\{\Fading[0], \Fading[1], \ldots\}$ can be exactly reconstructed (in the mean-square sense) by linearly combining the entire past $\{\Fading[-1], \Fading[-2], \ldots\}$. It has been established in \cite{lapidoth05:it} \cite{lapidoth03:it} that, for regular fading processes,
\begin{eqnarray}
\label{eqn:high-snr-regular}
C = \log\log\SNR - 1 - \gamma + \log\frac{1}{\ErrorVar_\mathrm{pred}} + o(1) \quad \mathrm{as}\; \SNR \rightarrow \infty,
\end{eqnarray}
where $\gamma = 0.5772\ldots$ is Euler's constant, and for deterministic fading processes,
\begin{eqnarray}
\label{eqn:high-snr-deterministic}
\frac{C}{\log\SNR} \rightarrow \frac{1}{2\pi}\cdot\mu\left(\left\{ \Omega: \spectral(e^{j\Omega}) = 0 \right\}\right) \quad \mathrm{as}\; \SNR \rightarrow \infty,
\end{eqnarray}
where $\mu(\cdot)$ denotes the Lebesgue measure on the interval $[-\pi, \pi]$.

It is then an interesting issue to investigate the connection between $(1/2\pi) \cdot \int_{-\pi}^\pi \spectral^2(e^{j\Omega})d\Omega$ and $\ErrorVar_\mathrm{pred}$. However, as the following two examples reveal, there is no explicit relationship between these two quantities.

\subsubsection{Example 1: Even a deterministic fading process can lead to poor low-SNR performance}
Consider the following class of spectral density functions $\spectral(e^{j\Omega})$ as illustrated in Figure \ref{fig:example1}:
\begin{figure}[tbp]
\psfrag{xlabel}{$\Omega$}
\psfrag{ylabel}{$\spectral(e^{j\Omega})$}
\epsfxsize=3.3in
\epsfclipon
\centerline{\epsffile{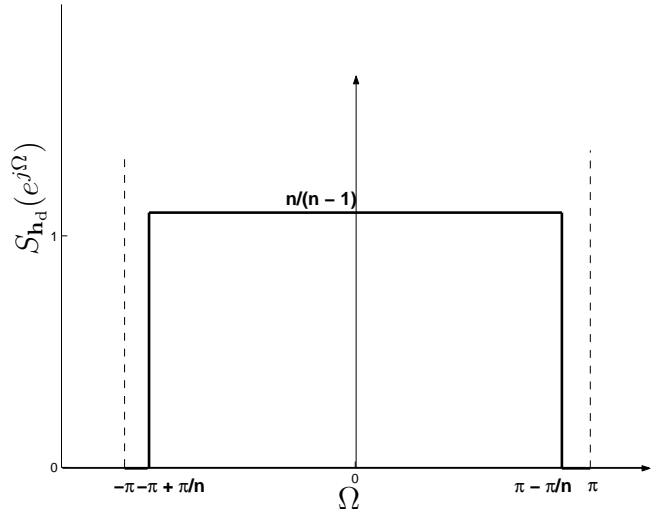}}
\caption{Spectral density function of a deterministic fading process that leads to poor low-SNR performance. The narrow notches on the spectrum make the process deterministic, while the remaining almost unit spectrum makes it behave as if nearly memoryless in the low-SNR regime for large $n$.}
\label{fig:example1}
\end{figure}
\begin{eqnarray*}
\spectral(e^{j\Omega}) = \left \{
\begin{array}{ll}
	\frac{n}{n - 1} & \mathrm{if}\;|\Omega| \leq \pi - \frac{\pi}{n}\\
	0 & \mathrm{if}\;\pi - \frac{\pi}{n} < |\Omega| \leq \pi
\end{array}
\right .,\quad n = 2, 3, \ldots
\end{eqnarray*}
Since $\spectral(e^{j\Omega}) = 0$ for certain intervals with non-zero measure, the corresponding fading process is deterministic with $\ErrorVar_\mathrm{pred} = 0$ \cite{doob53:book}. However, this class of $\spectral(e^{j\Omega})$ leads to
\begin{eqnarray*}
\frac{1}{2\pi}\int_{-\pi}^\pi \spectral^2(e^{j\Omega})d\Omega = \frac{n}{n - 1} \rightarrow 1\quad\mathrm{as}\;n \rightarrow \infty,
\end{eqnarray*}
resulting in vanishing values of the quadratic coefficient in (\ref{eqn:minfo-expand}).

\subsubsection{Example 2: Even an almost memoryless fading process can lead to good low-SNR performance}
Consider the following class of spectral density functions $\spectral(e^{j\Omega})$ as illustrated in Figure \ref{fig:example2}:
\begin{figure}[tbp]
\psfrag{xlabel}{$\Omega$}
\psfrag{ylabel}{$\spectral(e^{j\Omega})$}
\epsfxsize=3.3in
\epsfclipon
\centerline{\epsffile{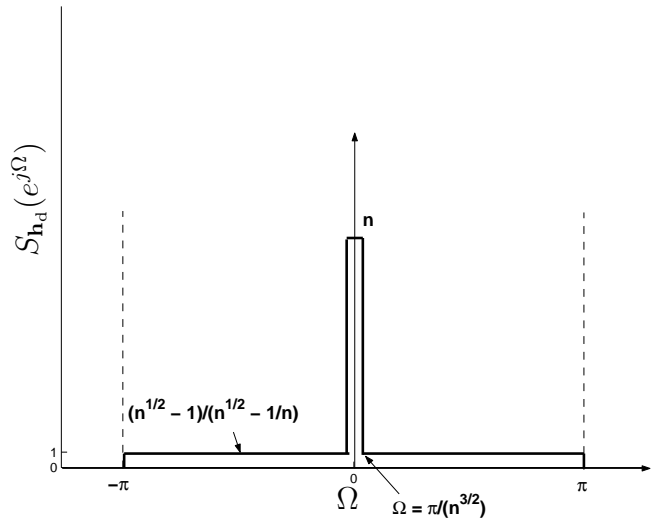}}
\caption{Spectral density function of an almost memoryless fading process that leads to good low-SNR performance. The almost unit spectrum makes the process nearly memoryless, while the narrow impulse-like spectrum peak significantly contributes to the integral $(1/2\pi)\cdot\int_{-\pi}^\pi \spectral^2(e^{j\Omega})d\Omega$, leading to good low-SNR performance for large $n$.}
\label{fig:example2}
\end{figure}
\begin{eqnarray*}
\spectral(e^{j\Omega}) = \left \{
\begin{array}{ll}
	n & \mathrm{if}\;|\Omega| \leq \frac{\pi}{n\sqrt{n}}\\
	\frac{\sqrt{n} - 1}{\sqrt{n} - 1/n} & \mathrm{if}\;\frac{\pi}{n\sqrt{n}} < |\Omega| \leq \pi
\end{array}
\right .,\quad n = 2, 3, \ldots
\end{eqnarray*}
For large $n$ the fading process becomes almost memoryless since
\begin{eqnarray*}
\ErrorVar_\mathrm{pred} = \exp\left\{ \frac{\log n}{n \sqrt{n}} + \log\frac{\sqrt{n} - 1}{\sqrt{n} - 1/n} \cdot (1 - \frac{1}{n\sqrt{n}}) \right\} \rightarrow 1\quad\mathrm{as}\;n \rightarrow \infty.
\end{eqnarray*}
However, this class of $\spectral(e^{j\Omega})$ also leads to
\begin{eqnarray*}
\frac{1}{2\pi}\int_{-\pi}^\pi \spectral^2(e^{j\Omega})d\Omega = \sqrt{n} + (1 - \frac{1}{n\sqrt{n}})\cdot\left ( \frac{\sqrt{n} - 1}{\sqrt{n} - 1/n} \right )^2 \rightarrow \infty
\end{eqnarray*}
as $n \rightarrow \infty$.

\subsection{Case Study: Discrete-Time Gauss-Markov Fading Processes}

In this subsection, we apply Theorem \ref{thm:minfo-expand} to analyze a specific class of discrete-time fading processes, namely, the discrete-time Gauss-Markov fading processes. The fading process in the channel model can be described by a first-order auto-regressive (AR) evolution equation of the form
\begin{eqnarray}
\label{eqn:gauss-markov}
\Fading[k + 1] = \sqrt{1 - \MSE} \cdot \Fading[k] + \sqrt{\MSE} \cdot \Innovation[k + 1],
\end{eqnarray}
where the innovation sequence $\{\Innovation[k]: -\infty < k < \infty\}$ consists of i.i.d. $\cGaussian(0, 1)$ random variables, and $\Innovation[k + 1]$ is independent of $\{\Fading[i]: -\infty < i \leq k\}$. The innovation rate $\MSE$ satisfies $0 < \MSE \leq 1$.

The spectral density function $\spectral(e^{j\Omega})$ for such a process is
\begin{eqnarray}
\label{eqn:gauss-markov-spectral}
\spectral(e^{j\Omega}) = \frac{\MSE}{(2 - \MSE) - 2 \sqrt{1 - \MSE} \cdot \cos\Omega},\quad -\pi \leq \Omega \leq \pi.
\end{eqnarray}
Hence
\begin{eqnarray*}
\frac{1}{2\pi}\int_{-\pi}^\pi \spectral^2(e^{j\Omega})d\Omega = \frac{\MSE^2}{2\pi}\int_{-\pi}^\pi \frac{1}{\left ( (2 - \MSE) - 2\sqrt{1 - \MSE}\cdot\cos\Omega \right )^2} d\Omega = \frac{2}{\MSE} - 1.
\end{eqnarray*}
Applying Theorem \ref{thm:minfo-expand}, we find that for the discrete-time Gauss-Markov fading model,
\begin{eqnarray}
R = \left(\frac{1}{\MSE} - 1\right)\cdot\SNR^2 + o(\SNR^2)\quad\mathrm{as}\;\SNR \rightarrow 0.
\end{eqnarray}
For practical systems in which the fading processes are underspread \cite{biglieri98:it}, the innovation rate $\MSE$ typically ranges from $1.8 \times 10^{-2}$ to $3 \times 10^{-7}$ \cite{etkin03:it}. So the $(1/2)\cdot\SNR^2 + o(\SNR^2)$ rate penalty of PSK with respect to optimal, peak-limited signaling may well be essentially negligible at low SNR.

\begin{figure}[tbp]
\psfrag{xlabel}{$\SNR$ (dB)}
\psfrag{ylabel}{$\SNR_\infty$ (dB)}
\epsfxsize=3.3in
\epsfclipon
\centerline{\epsffile{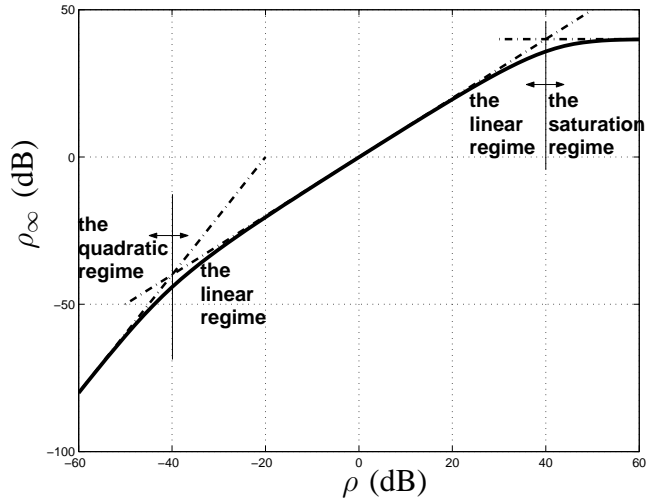}}
\caption{Case study of the discrete-time Gauss-Markov fading model: Illustration of the three operating regimes, $\MSE = 10^{-4}$.}
\label{fig:snr-inf}
\end{figure}

Due to the simplicity of the discrete-time Gauss-Markov fading model, we are able to carry out a non-asymptotic analysis to gain more insight. Applying (\ref{eqn:gauss-markov-spectral}) to (\ref{eqn:errorvar-inf}), the steady-state limiting channel prediction error is
\begin{eqnarray}
\label{eqn:gauss-markov-errorvar-inf}
\ErrorVar_\infty = \frac{(\SNR - 1)\cdot\MSE + \sqrt{(\SNR - 1)^2\cdot\MSE^2 + 4\SNR\MSE}}{2\SNR}.
\end{eqnarray}
Further applying (\ref{eqn:gauss-markov-errorvar-inf}) to (\ref{eqn:snr-inf}), we can identify the following three qualitatively distinct operating regimes of the induced channel (\ref{eqn:channel-coh}) for small $\MSE \ll 1$:
\begin{itemize}
\item The quadratic regime: For $\SNR \ll \MSE,\quad \ErrorVar_\infty \approx 1 - \SNR/\MSE,\quad \SNR_\infty \approx \SNR^2/\MSE$;
\item The linear regime: For $\MSE \ll \SNR \ll 1/\MSE,\quad \ErrorVar_\infty \approx \sqrt{\MSE/\SNR},\quad \SNR_\infty \approx \SNR$;
\item The saturation regime: For $1/\MSE \ll \SNR,\quad \ErrorVar_\infty \approx \MSE,\quad \SNR_\infty \approx 1/\MSE$.
\end{itemize}
Figure \ref{fig:snr-inf} illustrates these three regimes for $\MSE = 10^{-4}$. The different slopes of $\SNR_\infty$ on the log-log plot are clearly visible for the three regimes. The linear regime covers roughly $80$ dB, from $-40$ dB to $+40$ dB, in this particular example.

An interesting observation is that the two SNR thresholds dividing the three regimes are determined by a single parameter $\MSE$, which happens to be the one-step MMSE prediction error $\ErrorVar_\mathrm{pred}$ for the discrete-time Gauss-Markov fading process. The $1/\MSE$ threshold dividing the linear and the saturation regimes coincides with that obtained in \cite{etkin03:it}, where it is obtained for circular complex Gaussian inputs with nearest-neighbor decoding. In this paper we investigate PSK, which results in a penalty in the achievable rate at high SNR. More specifically, it can be shown that the achievable rate for $\SNR \gg 0$ behaves like $(1/2)\cdot\log \min\{\SNR, 1/\MSE\} + O(1)$ \cite{wyner66:bstj}.

\begin{figure}[tbp]
\psfrag{xlabel}{$\SNR$ (dB)}
\psfrag{ylabel}{$R/\SNR$}
\epsfxsize=3.3in
\epsfclipon
\centerline{\epsffile{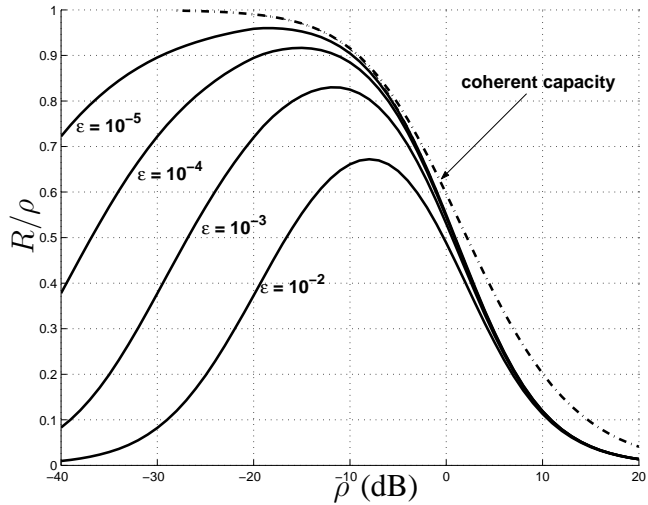}}
\caption{Normalized rate $R/\SNR$ vs. SNR for recursive training with QPSK on the discrete-time Gauss-Markov fading channel. As a comparison, the dashed-dot curve is the channel capacity (normalized by SNR) with perfect receive CSI, achieved by circular complex Gaussian inputs.}
\label{fig:gauss-markov-rate}
\end{figure}

A further observation relevant to low-SNR system design is that, the $\MSE$ threshold dividing the quadratic and the linear regimes clearly indicates when the low-SNR asymptotic channel behavior becomes dominant. Since the innovation rate $\MSE$ for underspread fading processes is typically small, we essentially have a low-SNR channel with perfect receive CSI above $\SNR = \MSE$. This suggests that there may be an ``optimal'' SNR at which the low-SNR capacity limit is the most closely approached. Figure \ref{fig:gauss-markov-rate} plots the normalized achievable rate $R/\SNR$ vs. SNR, in which the achievable rate $R$ is numerically evaluated for the induced channel (\ref{eqn:channel-coh}) using QPSK. Although all the curves vanish rapidly below the threshold $\SNR = \MSE$, for certain $\SNR > \MSE$, the normalized achievable rate can be reasonably close to $1$. For example, taking $\MSE = 10^{-4}$, the ``optimal'' SNR is $\SNR \approx -15$ dB, and the corresponding normalized achievable rate is above $0.9$, {\it i.e.}, more than $90\%$ of the low-SNR capacity limit is achieved.

\section{Filling the Gap to Capacity by Widening the Input Bandwidth}
\label{sec:bandwidth}

In Section \ref{sec:asymptotic} we have investigated the achievable information rate of the discrete-time channel (\ref{eqn:channel-dt}), which is obtained from the continuous-time channel (\ref{eqn:channel-ct}) as described in Section \ref{sec:model}. The symbol duration $T$ there is a fixed system parameter. In this section we will show that, if we are allowed to reduce $T$, {\it i.e.}, widen the input bandwidth, then the recursive training scheme using PSK achieves an information rate that is asymptotically consistent with the channel capacity under peak envelope $P$. More specifically, we have the following theorem.
\begin{thm}
\label{thm:minfo-ct}
For the continuous-time channel (\ref{eqn:channel-ct}) with envelope $P > 0$, as the symbol duration $T \rightarrow 0$, its induced channel (\ref{eqn:channel-coh}) achieves
\begin{eqnarray}
\label{eqn:minfo-ct}
\lim_{T \rightarrow 0} \frac{R}{T} = \left[1 - \frac{1}{P}\cdot\frac{1}{2\pi}\int_{-\infty}^\infty \log\left(1 + P \cdot \cspectral(j\omega)\right)d\omega\right]\cdot P.
\end{eqnarray}
\end{thm}
{\it Proof:} In Section \ref{sec:model} we have noted that the spectral density function $\spectral(e^{j\Omega})$ of the discrete-time fading process is related to $\cspectral(j\omega)$ through
\begin{eqnarray*}
\spectral(e^{j\Omega}) = \frac{1}{\sqrt{\int_0^T\int_0^T K_{\cFading}(s - t) ds dt}} \cdot \sum_{k = -\infty}^\infty \cspectral(j\frac{\Omega - 2k\pi}{T})\cdot \mathrm{sinc}^2(\Omega - 2k\pi),
\end{eqnarray*}
and that the SNR of the discrete-time channel (\ref{eqn:channel-dt}) is given by
\begin{eqnarray*}
\SNR = \left(\int_0^T\int_0^T K_{\cFading}(s - t) ds dt\right)\cdot\frac{P}{T}.
\end{eqnarray*}
For the proof, the following two identities are useful:
\begin{eqnarray}
\label{eqn:integral-zero-T}
\lim_{T \rightarrow 0} \frac{1}{T^2}\cdot\left(\int_0^T \int_0^T K_{\cFading}(s - t) dsdt\right) &=& 1\\
\label{eqn:ct-dt-spectral}
\lim_{T \rightarrow 0} \frac{1}{T}\cdot\frac{1}{2\pi}\int_{-\pi}^\pi \log\left( 1 + \SNR \cdot \spectral(e^{j\Omega}) \right)d\Omega &=& \frac{1}{2\pi}\int_{-\infty}^\infty \log\left( 1 + P\cdot \cspectral(j\omega) \right)d\omega.
\end{eqnarray}
The second identity (\ref{eqn:ct-dt-spectral}) has been established in \cite[Claims 8.1 and 8.2]{sethuraman05:it}. To prove the first one (\ref{eqn:integral-zero-T}), note that for the mean-square continuous fading process $\left\{\cFading(t): -\infty < t < \infty \right\}$, its autocorrelation function $K_{\cFading}(\tau) = \expect\{\cFading(t + \tau)\cFading\ctranspose(t)\}$ is continuous for all $-\infty < \tau < \infty$. Hence for any $T > 0$, there exists $T^\ast \in [0, T]$ such that
\begin{eqnarray*}
\int_0^T \int_0^T K_{\cFading}(s - t) dsdt &=& K_{\cFading}(T^\ast) \cdot T^2\\
&\rightarrow& K_{\cFading}(0) \cdot T^2 = T^2 \quad \mathrm{as}\;T \rightarrow 0.
\end{eqnarray*}
Now substituting (\ref{eqn:integral-zero-T}) and (\ref{eqn:ct-dt-spectral}) into (\ref{eqn:errorvar-inf}), we have
\begin{eqnarray}
\ErrorVar_\infty &=& \frac{1}{\SNR}\cdot \left\{\exp\left\{ \frac{1}{2\pi}\int_{-\pi}^\pi \log\left(1 + \SNR\cdot \spectral(e^{j\Omega})\right)d\Omega \right\} - 1\right\}\nonumber\\
&=& \frac{1}{\SNR}\cdot \left\{ \exp\left\{ \frac{1}{2\pi}\int_{-\infty}^\infty \log\left( 1 + P\cdot\cspectral(j\omega) \right)d\omega \cdot T + o(T) \right\} - 1 \right\}\nonumber\\
&=& \frac{1}{\SNR}\cdot \left[ \frac{1}{2\pi}\int_{-\infty}^\infty \log\left( 1 + P\cdot \cspectral(j\omega) \right)d\omega \cdot T + o(T) \right]\nonumber\\
&=& \frac{1}{P} \cdot \frac{1}{2\pi}\int_{-\infty}^\infty \log\left( 1 + P\cdot \cspectral(j\omega) \right)d\omega + o(1)\nonumber\quad\mathrm{as}\; T \rightarrow 0.
\end{eqnarray}
That is,
\begin{eqnarray}
\label{eqn:errorvar-ct-asymptotic}
\lim_{T \rightarrow 0} \ErrorVar_\infty = \frac{1}{P} \cdot \frac{1}{2\pi}\int_{-\infty}^\infty \log\left( 1 + P\cdot \cspectral(j\omega) \right)d\omega.
\end{eqnarray}
Then substituting (\ref{eqn:integral-zero-T}) and (\ref{eqn:errorvar-ct-asymptotic}) into (\ref{eqn:snr-inf}), we have
\begin{eqnarray}
\lim_{T \rightarrow 0} \frac{\SNR_\infty}{T} &=& \lim_{T \rightarrow 0} \frac{(1 - \ErrorVar_\infty)\cdot\SNR}{(\ErrorVar_\infty \cdot \SNR + 1)\cdot T}\nonumber\\
\label{eqn:snr-inf-lim}
&=& \left[1 - \frac{1}{P}\cdot\frac{1}{2\pi}\int_{-\infty}^\infty \log\left(1 + P \cdot \cspectral(j\omega)\right)d\omega\right]\cdot P.
\end{eqnarray}
Finally Theorem \ref{thm:minfo-ct} immediately follows from substituting (\ref{eqn:snr-inf-lim}) into (\ref{eqn:rate}).   {\bf Q.E.D.}

Again we compare the asymptotic achievable rate (\ref{eqn:minfo-ct}) to a capacity upper bound based upon the capacity per unit energy. For the continuous-time channel (\ref{eqn:channel-ct}), the capacity per unit energy under a peak envelope constraint $P > 0$ is \cite{sethuraman05:it}
\begin{eqnarray}
\label{eqn:capacity-ct}
\dot{C} = 1 - \frac{1}{2\pi P} \cdot \int_{-\infty}^\infty \log\left( 1 + P\cdot\cspectral(j\omega) \right)d\omega,
\end{eqnarray}
and the related capacity upper bound (measured per unit time) is \cite{sethuraman05:it}
\begin{eqnarray}
\label{eqn:capacity-ub-ct}
C \leq U(P) &\define& \left[1 - \frac{1}{2\pi P} \cdot \int_{-\infty}^\infty \log\left( 1 + P\cdot\cspectral(j\omega) \right)d\omega\right] \cdot P.
\end{eqnarray}
Comparing (\ref{eqn:minfo-ct}) and (\ref{eqn:capacity-ub-ct}), it is surprising to notice that these two quantities coincide. Recalling that in Section \ref{sec:asymptotic} we have noticed a $(1/2)\cdot\SNR^2 + o(\SNR^2)$ rate penalty in discrete-time channels, we conclude that widening the input bandwidth eliminates this penalty and essentially results in an asymptotically capacity-achieving scheme in the wideband regime\footnote{Again, the same caveat as in footnote 2 applies.}.

The channel capacity of continuous-time peak-limited wideband fading channels (\ref{eqn:minfo-ct}) was originally obtained in \cite{viterbi67:it}. However, in \cite{viterbi67:it} the capacity is achieved by frequency-shift keying (FSK), which is bursty in frequency. In our Theorem \ref{thm:minfo-ct}, we show that the capacity is also achievable if we employ recursive training and PSK, which is bursty in neither time nor frequency.

After some manipulations of (\ref{eqn:minfo-ct}), we further have that
\begin{itemize}
\item As $P \rightarrow 0$,
\begin{eqnarray}
\label{eqn:minfo-ct-smallP}
\frac{\lim_{T \rightarrow 0} (R/T)}{P^2} \rightarrow \frac{1}{2}\cdot\frac{1}{2\pi}\int_{-\infty}^\infty \cspectral^2(j\omega)d\omega,
\end{eqnarray}
if the above integral exists.
\item As $P \rightarrow \infty$,
\begin{eqnarray}
\label{eqn:minfo-ct-largeP}
\frac{\lim_{T \rightarrow 0} (R/T)}{P} \rightarrow 1.
\end{eqnarray}
\end{itemize}
In the sequel we will see that (\ref{eqn:minfo-ct-smallP}) and (\ref{eqn:minfo-ct-largeP}) are useful for asymptotic analysis.

\subsection{An Intuitive Explanation of Theorem \ref{thm:minfo-ct}}
In our proof of Theorem \ref{thm:minfo-ct}, we have utilized identities (\ref{eqn:integral-zero-T}) and (\ref{eqn:ct-dt-spectral}) to conveniently relate the continuous-time channel (\ref{eqn:channel-ct}) to the discrete-time channel (\ref{eqn:channel-dt}). However, these identities also have concealed much of the intuition contained in the derivation. To further illustrate the underlying mechanism in Theorem \ref{thm:minfo-ct}, here we give an alternative argument. Although the following reasoning is not mathematically rigorous, it does provide an intuitive way to understand the channel behavior as the symbol duration $T \rightarrow 0$.

In Section \ref{sec:model}, we have described the conversion from the continuous-time channel (\ref{eqn:channel-ct}) to the discrete-time channel (\ref{eqn:channel-dt}). Strictly speaking, the discrete-time fading coefficient $\Fading[k]$ is the $k$th sample of the matched-filtered, continuous-time fading process. The matched-filtering effect can be viewed as averaging $\cFading(t)$ within a symbol interval of length $T$. Since we have assumed that the continuous-time fading process $\{\cFading(t): -\infty < t < \infty\}$ is mean-square continuous in $t$, roughly speaking, as $T \rightarrow 0$, the discrete-time fading coefficient $\Fading[k] \approx \cFading(kT)$, and the SNR per symbol $\SNR \approx P\cdot T$. Furthermore, compared to sufficiently small $T$, the fading process $\{\cFading(t): -\infty < t < \infty\}$ can be viewed as essentially band-limited. So the discrete-time fading process $\{\Fading[k]: -\infty < k < \infty\}$ is approximately the sampled continuous-time fading process $\{\cFading(t): -\infty < t < \infty\}$ with sampling rate well beyond its Nyquist rate, and we may write $\spectral(e^{j\Omega}) \approx (1/T)\cdot \cspectral(j\Omega/T)$ for $-\pi \leq \Omega \leq \pi$.

Now let us apply the above approximations to (\ref{eqn:errorvar-inf}) to evaluate $\ErrorVar_\infty$ for small $T$:
\begin{eqnarray*}
\ErrorVar_\infty &=& \frac{1}{\SNR} \cdot\left\{ \exp\left\{ \frac{1}{2\pi}\int_{-\pi}^\pi \log\left( 1 + \SNR\cdot\spectral(e^{j\Omega}) \right) d\Omega \right\} - 1 \right\}\\
&\approx& \frac{1}{P\cdot T} \cdot\left\{ \exp\left\{ \frac{1}{2\pi}\int_{-\pi}^\pi \log\left( 1 + P\cdot\cspectral(j\frac{\Omega}{T}) \right) d\Omega \right\} - 1 \right\}\\
&=& \frac{1}{P\cdot T} \cdot\left\{ \exp\left\{ \frac{1}{2\pi}\int_{-\pi/T}^{\pi/T} \log\left( 1 + P\cdot\cspectral(j\Omega) \right) d\Omega \cdot T \right\} - 1 \right\}\\
&\approx& \frac{1}{P\cdot T} \cdot\left\{ \exp\left\{ \frac{1}{2\pi}\int_{-\infty}^\infty \log\left( 1 + P\cdot\cspectral(j\omega) \right) d\omega \cdot T \right\} - 1 \right\}\\
&\approx& \frac{1}{P\cdot T} \cdot \frac{1}{2\pi}\int_{-\infty}^\infty \log\left( 1 + P\cdot\cspectral(j\omega) \right)d\omega \cdot T = \frac{1}{P} \cdot \frac{1}{2\pi}\int_{-\infty}^\infty \log\left( 1 + P\cdot\cspectral(j\omega) \right)d\omega,
\end{eqnarray*}
which is the same as (\ref{eqn:errorvar-ct-asymptotic}) in our proof.

\subsection{Case Study: The Continuous-Time Gauss-Markov Fading Model}

\begin{figure}[tbp]
\psfrag{xlabel}{$P$ (dB)}
\psfrag{ylabel}{$\lim_{T \rightarrow 0} (R/T)$}
\epsfxsize=3.3in
\epsfclipon
\centerline{\epsffile{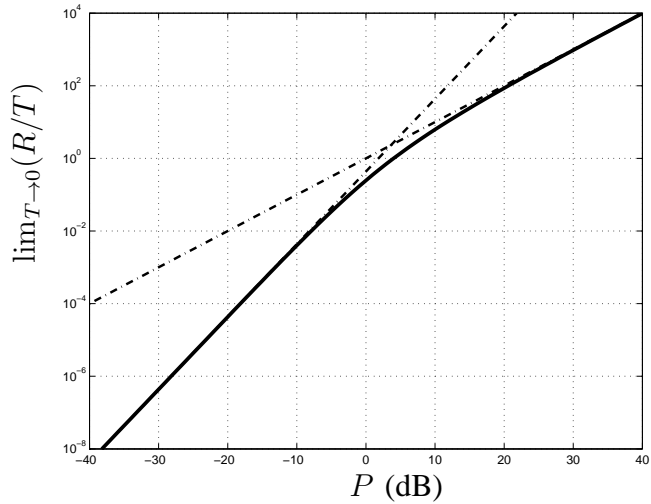}}
\caption{The asymptotic achievable rate $\lim_{T \rightarrow 0} (R/T)$ vs. the envelope $P$, for recursive training with complex proper PSK on the continuous-time Gauss-Markov fading channel with innovation rate $\epsilon_\mathrm{c} = 0.9$. The dashed-dot curves indicate the limiting behaviors for small and large $P$.}
\label{fig:ct-gauss-markov}
\end{figure}

In this subsection, we apply Theorem \ref{thm:minfo-ct} to analyze the continuous-time Gauss-Markov fading processes. Such a process has autocorrelation function
\begin{eqnarray*}
K_{\cFading}(\tau) = (1 - \epsilon_\mathrm{c})^{|\tau|/2},
\end{eqnarray*}
where the parameter $0 < \epsilon_\mathrm{c} \leq 1$ characterizes the channel variation, analogously to $\epsilon$ for the discrete-time case in Section \ref{sec:asymptotic}. The spectral density function of the process is
\begin{eqnarray*}
\cspectral(j\omega) = \frac{|\log (1 - \epsilon_\mathrm{c})|}{\omega^2 + \left(\log (1 - \epsilon_\mathrm{c})\right)^2/4}.
\end{eqnarray*}
Applying Theorem \ref{thm:minfo-ct}, we find that the recursive training scheme using PSK with a wide bandwidth asymptotically achieves an information rate
\begin{eqnarray}
\label{eqn:minfo-gm-ct}
\lim_{T \rightarrow 0} \frac{R}{T} &=& P - \frac{|\log(1 - \epsilon_\mathrm{c})|}{2}\cdot\left( \sqrt{1 + \frac{4P}{|\log(1 - \epsilon_\mathrm{c})|}} - 1 \right)\\
&=& \frac{1}{|\log(1 - \epsilon_\mathrm{c})|}\cdot P^2 + o(P^2)\quad \mathrm{as}\; P \rightarrow 0.
\end{eqnarray}
Figure \ref{fig:ct-gauss-markov} illustrates the achievable rate (\ref{eqn:minfo-gm-ct}) vs. $P$ for $\epsilon_\mathrm{c} = 0.9$.

\subsection{Case Study: Clarke's Fading Model}

In this subsection, we apply Theorem \ref{thm:minfo-ct} to analyze Clarke's fading processes. Such a process is usually characterized by its spectral density function \cite{jakes94:book}
\begin{eqnarray*}
\cspectral(j\omega) = \left\{
\begin{array}{ll}
	\frac{2}{\omega_m}\cdot\frac{1}{\sqrt{1 - (\omega/\omega_m)^2}}, & \mathrm{if}\;|\omega| \leq \omega_m\\
	0, & \mathrm{otherwise},
\end{array}
\right .
\end{eqnarray*}
where $\omega_m$ is the maximum Doppler frequency.

\begin{figure}[tbp]
\psfrag{xlabel}{$P$ (dB)}
\psfrag{ylabel}{$\lim_{T \rightarrow 0} (R/T)$}
\epsfxsize=3.3in
\epsfclipon
\centerline{\epsffile{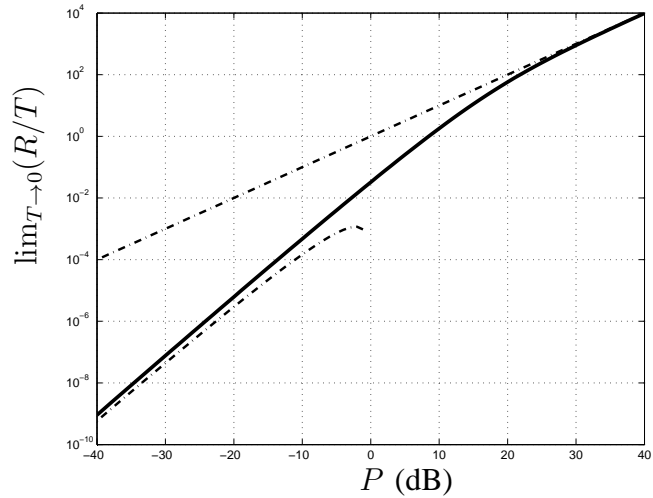}}
\caption{The asymptotic achievable rate $\lim_{T \rightarrow 0} (R/T)$ vs. the envelope $P$, for recursive training with complex proper PSK on Clarke's fading channel with maximum Doppler frequency $\omega_m = 100$. The dashed-dot curves indicate the limiting behaviors for small and large $P$.}
\label{fig:clarkes}
\end{figure}

Applying Theorem \ref{thm:minfo-ct}, we find that
\begin{eqnarray}
\label{eqn:minfo-clarkes-ct}
\lim_{T \rightarrow 0} \frac{R}{T} = \left\{
\begin{array}{ll}
	\frac{\omega_m}{\pi}\cdot\left\{ \log\frac{\omega_m}{P} - \sqrt{1 - (2P/\omega_m)^2}\cdot\log\frac{\omega_m \cdot\left[1 + \sqrt{1 - (2P/\omega_m)^2}\right]}{2P} \right\}, & \mathrm{if}\; P \leq \omega_m/2\\
	\frac{\omega_m}{\pi}\cdot\left\{ \log\frac{\omega_m}{P} + \sqrt{(2P/\omega_m)^2 - 1}\cdot \arctan \sqrt{(2P/\omega_m)^2 - 1} \right\}, & \mathrm{if}\; P > \omega_m/2.
\end{array}
\right .
\end{eqnarray}
For large $P$, the asymptotic behavior of (\ref{eqn:minfo-clarkes-ct}) is consistent with (\ref{eqn:minfo-ct-largeP}). For small $P$, however, the integral in (\ref{eqn:minfo-ct-smallP}) diverges, hence the asymptotic behavior of (\ref{eqn:minfo-clarkes-ct}) scales super-quadratically with $P$. After some manipulations of (\ref{eqn:minfo-clarkes-ct}), we find that
\begin{eqnarray}
\label{eqn:minfo-clarkes-smallP}
\lim_{T \rightarrow 0} \frac{R}{T} = \frac{2}{\pi \omega_m}\cdot \log\frac{1}{P}\cdot P^2 + O(P^2)\quad\mathrm{as}\;P \rightarrow 0.
\end{eqnarray}
Figure \ref{fig:clarkes} illustrates the achievable rate (\ref{eqn:minfo-clarkes-ct}) vs. $P$ for $\omega_m = 100$. We notice that, for small $P$, the asymptotic expansion (\ref{eqn:minfo-clarkes-smallP}) is accurate.

\section{Concluding Remarks}
\label{sec:conclusion}

For fading channels that exhibit temporal correlation, a key to enhancing communication performance is efficiently exploiting the implicit CSI embedded in the fading processes. From the preceding developments in this paper, we see that a recursive training scheme, which performs channel estimation and demodulation/decoding in an alternating manner, accomplishes this job reasonably well, especially when the channel fading varies slowly. The main idea of recursive training is to repeatedly use decisions of previous information symbols as pilots, and to ensure the reliability of these decisions by coding over sufficiently long blocks.

Throughout this paper, we restrict the channel inputs to complex proper PSK, which is not optimal in general for Rayleigh fading channels without CSI. There are two main motivations for this choice. First, compared to other channel inputs such as circular complex Gaussian, PSK leads to a significant simplification of the analytical developments. As we saw, recursive training with PSK converts the original fading channel without CSI into a series of parallel sub-channels, each with estimated receive CSI but additional noise that remains circular complex white Gaussian. In this paper we mainly investigate the steady-state limiting channel behavior; however, it may worth mentioning that, using the induced channel model presented in Section \ref{sec:model}, exact evaluation of the transient channel behavior is straightforward, with the aid of numerical methods.

Second, PSK inputs perform reasonably well in the moderate to low SNR regime. This is due to the fact that, for fading channels with perfect receive CSI, as SNR vanishes, channel capacity can be asymptotically achieved by rather general complex proper inputs besides circular complex Gaussian \cite{prelov04:it}. The main contribution of our work is that it clearly separates the effect of an input peak-power constraint and the effect of replacing optimal peak-limited inputs with PSK, which is non-bursty in both time and frequency. It is shown that, for slowly time-varying fading processes, the rate loss from PSK inputs is essentially negligible. Furthermore, as revealed by the non-asymptotic analysis for discrete-time Gauss-Markov fading processes, there appear to be non-vanishing SNRs at which near-coherent performance is attainable with recursive training and PSK.

\section*{Acknowledgment}
The authors are grateful to Shlomo Shamai (Shitz) for bringing to attention reference \cite{viterbi67:it}, and for providing useful comments on this work.

\bibliographystyle{ieee}
\bibliography{../proposal/prop}

\end{document}